\documentclass
[aps,12pt,a4paper,amsfonts,amssymb,titlepage,showpacs,preprint,showkeys,nofootinbib,nobibnotes,byrevtex,superscriptaddress]{revtex4}%
\usepackage{amsfonts}
\usepackage{amsmath}
\usepackage{amssymb}
\usepackage{graphicx}%
\providecommand{\U}[1]{\protect\rule{.1in}{.1in}}

\begin{document}
\title{The Relation Between Gauge and Non-Gauge Abelian Models}
\author{Gabriel Di Lemos Santiago Lima}
\affiliation{Institute for Gravitation and the Cosmos, and Physics Department, The Pennsylvania State University, University Park, PA 16802, USA}
\affiliation{Centro Federal de Educa\c{c}\~{a}o Tecnol\'ogica Celso Suckow da Fonsceca, Uned Nova Igua\c{c}u, Estrada de Adrian\'{o}polis, 1317, Santa Rita, Nova Igua\c{c}u, RJ, CEP 26041-271, BR}
\email{gabriel@gravity.psu.edu, gabriellemos3@hotmail.com}

\begin{abstract}
This work studies the relationship between gauge-invariant and non gauge-invariant abelian vector models. Following a technique introduced by Harada and Tsutsui, we show that the Proca and the Chiral Schwinger models may both be viewed as gauge-fixed versions of genuinely gauge-invariant models. This leads to the proposal that any consistent Abelian vector model with no gauge symmetry can be understood as a gauge theory that had its gauge fixed, which establishes an equivalence between gauge-invariant and non gauge-invariant models. Finally, we show that a gauge-invariant version of the chiral Schwinger model, after integrating out the fermionic degrees of freedom, can be identified with the two-dimensional Stueckelberg model without the gauge fixing term.
\end{abstract}
\maketitle

\section{Introduction}

It is well known that anomalous gauge theories usually spoil unitarity and renormalizability due to the breakdown of gauge symmetry at the quantum level \cite{BIM, GJ}. Yet, it is also believed that the gauge anomaly breaks the current conservation law. In view of this, gauge anomalous models are usually considered as being inconsistent.

Contrary to this idea, a group of authors has shown that gauge symmetry may be restored from anomalous models by the addition of extra degrees of freedom. Indeed the work of Fadeev and Shatashvilli \cite{FS} restores the gauge symmetry of the final effective action by adding a Wess-Zumino term to the fundamental action. Soon after, the works of Babelon, Shaposhnik and Viallet \cite{BSV} and Harada and Tsutsui \cite{HT} showed, independently, that such a Wess-Zumino term could be derived through algebraic manipulations over the functional integral. Then, it became clear that such a way of deriving gauge-invariant models from anomalous ones did not need to be limited to the case of anomalous models, but it could also be applied to non anomalous ones that do not exhibit classical gauge symmetry since the beginning, like the Proca Model, also analyzed by the last cited authors \cite{HT2}, thus, leading to a natural generalization of their technique. This naturally conducts us to the conclusion that anomalous models are analogous to \textit{any} theory that has no gauge symmetry and, thus, may be treated in the same way, in order to restore gauge invariance.  

To understand the role played by the emerging extra field in the Abelian case, a recent work has shown that the gauge-invariant formulation of the Proca model may be identified with the Stueckelberg theory \cite{Stueck}, leading to the interpretation of such a field as being the Stueckelberg scalar \cite{eu}.

At this point, we may ask, in general, if both formulations (the gauge-invariant, and the non invariant one) can be taken as being physically equivalent and, in particular, whether the current of both is conserved or not. Yet, only in the context of anomalous models, the Harada-Tsutsui technique (HT) leads to two distinct ways of achieving the same gauge-invariant effective theory, \textit{before} the integration over the matter fields: the one which adds up the Wess-Zumino term, known as the \textit{standard} formulation \cite{HT}, and another one which also couples the extra degrees of freedom with the matter fields, called the \textit{enhanced} formulation \cite{eu}. In this sense, we may ask whether both ways are redundant or if the informations contained in each of them are physically distinguishable.

This work is intended to elucidate these questions for the case of Abelian vector models, and the relation between original Abelian anomalous models, the standard formulation and the enhanced one is analyzed, as well as the relation between the Proca and Stueckelberg's models. In this sense, in section I, the \textit{enhanced} version of Harada-Tsutsui gauge-invariant mapping is derived, as well as the \textit{standard} one. In section II, we rederive the Stueckelberg model from the Proca one, and the analysis of both formulations shows their equivalence. In section III, the same kind of analysis is done, but comparing the enhanced version of general Abelian anomalous gauge models with the original ones. It is shown that if we alternatively consider that the current is conserved by the equation of motion of the gauge field, as an analogue to the subsidiary condition arising in the Proca model, then both formulations become equivalent, since the first may be reduced to the second by a gauge condition which represents the anomaly cancelation of the original model. The chiral Schwinger model is used as an example. It is also shown, in more details, that the enhanced formulation of Abelian anomalous models is free from anomalies.  

Then, the two examples analysed lead us, naturally, to an equivalence statement related to gauge and non-gauge Abelian models, which is done in section IV. Yet in this section, it is shown that if the anomaly is not gauge-invariant, it still remains in the standard formulation, and that this one may be equivalent to the other formulations only if the anomaly is gauge-invariant. Finally, in section V, it is shown that, after integrating out the fermions, a gauge-invariant formulation of the chiral Schwinger model may be identified with the original Stueckelberg model in two dimensions. We, thus, conclude this work in section VI.

\section{Enhanced version of Harada-Tsutsui gauge-invariant procedure}

We consider an \textit{anomalous} generic Abelian effective action, defined by

\begin{equation}
\exp(iW[A_\mu])=\int d\psi d\bar{\psi}\exp(iI[\psi, \bar{\psi}, A_\mu]),
\label{effective} 
\end{equation}
where $I[\psi, \bar{\psi}, A_\mu]$ is invariant under local gauge transformations

\begin{align}
A_{\mu}^{\theta}&=A_\mu+\frac{1}{e}\partial_\mu\theta(x),\\
\psi&=\exp(i\theta(x))\psi,\\
\bar{\psi}&=\exp(-i\theta(x))\bar{\psi},
\end{align}
that is,

\begin{equation}
I\left[\psi^{\theta},\bar{\psi}^\theta,A^{\theta}\right]=I\left[\psi,\bar{\psi},A\right],
\end{equation}
 while, by definition,
 
 \begin{equation}
 W\left[A^{\theta}_{\mu}\right]\neq W\left[A_{\mu}\right]. \label{noninv_effect_act}
 \end{equation}
 The formulation with the addition of the Wess-Zumino term, first proposed by Fadeev and Shatashvilli \cite{FS}, and then derived by Harada and Tsutsui \cite{HT}, arises when we go to the full quantum theory by redefining the vacuum functional
 
 \begin{equation}
 Z=\int dA_\mu d\psi d\bar{\psi}\exp(iI[\psi,\bar{\psi},A_{\mu}])=\int dA_{\mu}\exp(iW[A_{\mu}])
 \end{equation}
 multiplying it by the gauge volume
 
  \begin{equation}
 Z=\int d\theta dA_\mu d\psi d\bar{\psi}\exp(iI[\psi,\bar{\psi},A_{\mu}])=\int d\theta dA_{\mu}\exp(iW[A_{\mu}]).
 \end{equation}
 We, then, change variables in the gauge field so that
 
 \begin{equation}
 A_{\mu}\rightarrow A_{\mu}^{\theta},\qquad\qquad dA_{\mu}\rightarrow dA_{\mu}^{\theta},
 \end{equation}
 and use translational invariance of $dA_{\mu}$, so that
 
 \begin{equation}
 dA_{\mu}^{\theta}=dA_{\mu},
 \end{equation}
 to reach the final gauge-invariant effective action, which takes the field $\theta$ into account, defined by
 
 \begin{equation}
 \exp(iW_{eff}[A_{\mu}])\equiv\int d\theta \exp(iW[A_{\mu}^{\theta}]).
 \label{inv-effect}
 \end{equation} 
 Using $(\ref{effective})$, it is evident that
 
 \begin{equation}
\exp(iW[A_{\mu}^{\theta}])=\int d\psi d\bar{\psi} \exp(iI_{st}[\psi, \bar{\psi}, A_{\mu}, \theta]),
\end{equation}
where

\begin{equation}
I_{st}[\psi, \bar{\psi}, A_{\mu}, \theta]\equiv I[\psi, \bar{\psi}, A_{\mu}]+\alpha_{1}[A_{\mu}, \theta]
\label{standard}
\end{equation}
is called the \textit{standard} action, and

\begin{equation}
\alpha_{1}[A_{\mu}, \theta]\equiv W[A_{\mu}^{\theta}]-W[A_{\mu}]
\end{equation}
is known as the Wess-Zumino term \cite{HF}. It can be seen that, although the final effective action is gauge-invariant, the initial one $(\ref{standard})$ is not, since the Wess-Zumino term breaks gauge symmetry. On the other hand, we may raise an alternative gauge-invariant initial action by noticing that $(\ref{inv-effect})$ can also be obtained by

\begin{equation}
\exp(iW[A_{\mu}^{\theta}])=\int d\psi d\bar{\psi} \exp(iI_{en}[\psi, \bar{\psi}, A_{\mu}, \theta]),
\label{enh_act_funct}
\end{equation}
where

\begin{equation}
I_{en}[\psi, \bar{\psi}, A_{\mu}, \theta]\equiv I[\psi, \bar{\psi}, A_{\mu}^{\theta}].
\label{enh_act}
\end{equation}
This simplifies and systematizes the Harada-Tsutsui procedure by noticing that we only need to make the substitution $A_{\mu}\rightarrow A_\mu^{\theta}$ in the fundamental action, as it becomes clear in the example of the massive vector theory. It is also evident that, to obtain such a gauge-invariant formulation, we do not even need to proceed to such a substitution in the entire action. Indeed, we only need to add a gradient of a scalar to the gauge field in the parts of the initial action that do not remain gauge-invariant \textit{after} integrating out the fermions.

The inclusion of the field $\theta$ in the enhanced formulation also transforms it into a modified gauge theory, even before the integration over the scalar. To see this, we notice that such a formulation is invariant under Pauli's transformations \cite{Pauli}

\begin{eqnarray}
A_{\mu} &\rightarrow & A_{\mu}+\frac{1}{e}\partial_{\mu}\Lambda \nonumber\\
\theta &\rightarrow & \theta-\Lambda.
\end{eqnarray}
This was also noticed by Hadara and Tsutsui in the massive vector case \cite{HT2}. We shall distinguish between the scalar provided by the standard action and the one associated to the enhanced formulation, calling the first one a Wess-Zumino field, and the second one a Stueckelberg field.

\section{Equivalence Between the Proca and the Stueckelberg models}

Consider a Proca field interacting with fermions, which is described by the action

\begin{equation}
I_{P}[\psi, \bar{\psi}, A_{\mu}]\equiv I_{M}[\psi, \bar{\psi}, A_{\mu}]+W_{P}[A_{\mu}],
\end{equation}
where $I_{M}[\psi, \bar{\psi}, A_{\mu}]$ is the matter action minimally coupled to the abelian field $A_{\mu}$, that exhibits local gauge symmetry, and $W_{P}[A_{\mu}]$ is the pure Proca action, defined by

\begin{equation}
W_{P}[A]\equiv\int d^{n}x\left(-\frac{1}{4}F^{\mu\nu}F_{\mu\nu}+\frac {m^{2}}{2}A^{\mu}A_{\mu}\right). \nonumber
\end{equation}
Evidently, the action above has no gauge symmetry, since the massive term spoils it. The classical equations of motion are given by
\begin{eqnarray}
\frac{\delta I_{M}}{\delta\psi}&=&\frac{\delta I_{M}}{\delta\bar{\psi}}=0\\
\partial_{\mu}F^{\mu\nu}+m^{2}A^{\nu}&=&eJ^{\nu}, \label{motion-eq.A}
\end{eqnarray}
where
\begin{equation}
J^{\mu}=-\frac{1}{e}\frac{\delta I_{M}}{\delta A^{\mu}}
\end{equation}
is the conserved matter current obtained by global gauge invariance. If we now take the divergence of eq. (\ref{motion-eq.A}), we just arrive at
\begin{equation}
\partial_{\mu}A^{\mu}=0
\end{equation}
as a \textit{subsidiary} condition.

On the other hand, we could apply the HT technique by gauge transforming only the massive part of the action to obtain

\begin{equation}
I_{P(en)}[\psi, \bar{\psi}, A, \theta]=I_{M}[\psi, \bar{\psi}, A]+W_{P(en)}[A, \theta], \label{enh-Proca}
\end{equation}
where $W_{P(en)}[A, \theta]$ is just the pure enhanced Proca action, given by

\begin{equation}
W_{P(en)}[A,\theta]\equiv W_{P}[A^{\theta}]=-\frac{1}{4}\int d^{4}x F^{\mu\nu}F_{\mu\nu}+\frac{m^{2}}{2}\int d^{4}x\left(A^{\mu}+\frac{1}{e}\partial^{\mu}\theta\right)\left(A_{\mu}+\frac{1}{e}\partial_{\mu}\theta\right). \label{Stueckelberg}
\end{equation}
It is easy to notice that (\ref{Stueckelberg}) is just the Stueckelberg action. To see this, we notice that if we redefine the field $\theta$ as
\begin{equation}
B(x)\equiv \frac{m}{e}\theta (x),
\end{equation}
then, (\ref{Stueckelberg}) becomes exactly the Stueckelberg action \cite{Stueck}:
\begin{equation}
W_{Stueck}[A,B]=-\frac{1}{4}\int d^{4}(x)F^{\mu\nu}F_{\mu\nu}+\frac{1}{2}\int d^{4}x(mA^{\mu}+\partial^{\mu}B)(mA_{\mu}+\partial_{\mu}B).
\end{equation}

It is clear that the Stueckelberg model is reducible to the original Proca one by the gauge choice where the Stueckelberg field is set to be constant. But the result of interest for us is to show the equivalence between the Proca model and it's gauge-invariant version \textit{after} integrating out the field $\theta$. To this end, we may integrate the exponential of (\ref{Stueckelberg}) over the gauge orbits in order to find the gauge-invariant version of the Proca model coupled to the fermions:
\begin{equation}
\exp\left(iI_{P}\left[\psi, \bar{\psi},A\right]\right)\equiv \exp \left(iI_{M}\left[\psi,\bar{\psi},A\right]\right)\int d\theta \exp (iW_{P(en)}[A,\theta]).
\end{equation}
To do this, we notice that
\begin{equation}
\int d\theta \exp(iW_{P(en)}[A,\theta])=\exp(iW_{P}[A])\int d\theta \exp\left(i\int\frac{1}{2}\frac{m^{2}}{e^{2}}\partial^{\mu}\theta\partial_{\mu}\theta+\frac{m^2}{e}A^{\mu}\partial_{\mu}\theta\right),
\end{equation}
and that
\begin{align}
&\int d\theta \exp\left(i\int\frac{1}{2}\frac{m^{2}}{e^{2}}\partial^{\mu}\theta\partial_{\mu}\theta+\frac{m^2}{e}A^{\mu}\partial_{\mu}\theta\right)\nonumber\\
&=\exp\left(-\frac{i}{2}m^2\int d^{n}x A_{\mu}\frac{\partial^{\mu}\partial^{\nu}}{\square}A_{\nu}\right)\int d\theta \exp\left(-i\frac{m^{2}}{2e}\int d^{n}x\left[\left(\frac{e}{\square}\partial^{\mu}A_{\mu}+\theta\right)\square\left(\frac{e}{\square}\partial^{\nu}A_{\nu}+\theta\right)\right]\right).
\end{align}
Doing the following change of variables in the field $\theta$:
\begin{equation}
\theta\rightarrow\theta '=\theta+\frac{e}{\square}\partial^{\mu}A_{\mu},
\label{PC}
\end{equation}
we find
\begin{equation}
\int d\theta\exp\left(i\int\frac{1}{2}\frac{m^{2}}{e^{2}}\partial^{\mu}\theta\partial_{\mu}\theta+\frac{m^{2}}{e}A^{\mu}\partial_{\mu}\theta\right)\sim\exp\left(-\frac{i}{2}m^{2}\int d^{n}xA_{\mu}\frac{\partial^{\mu}\partial{\nu}}{\square}A_{\nu}\right),
\end{equation}
and,thus
\begin{equation}
I'_{P}[\psi,\bar{\psi},A]=I_{M}[\psi,\bar{\psi},A]+\int d^{n}\left\{-\frac{1}{4}F^{\mu\nu}F_{\mu\nu}+\frac{1}{2}m^{2}A_{\mu}\left(\eta^{\mu\nu}-\frac{\partial^{\mu}\partial^{\nu}}{\square}A_{\nu}\right)\right\}. \label{inv-Proca}
\end{equation}
Although we went far away going to the full quantum model to derive (\ref{inv-Proca}), we now use it classically and derive the equations of motion. Then, we just obtain
\begin{eqnarray}
\frac{\delta I_{M}}{\delta\psi}&=&\frac{\delta I_{M}}{\delta\bar{\psi}}=0\\
eJ^{\nu}&=&\partial_{\mu}F^{\mu\nu}+m^{2}\left(\eta^{\mu\nu}-\frac{\partial^{\mu}\partial^{\nu}}{\square}\right)A_{\nu},
\end{eqnarray}
and it becomes clear that the equations of motion of this gauge-invariant version of the massive vector model coincides with the Proca one if we fix the Lorenz gauge $\partial_{\mu}A^{\mu}=0$, showing the equivalence between both formulations. It can be seen that such a gauge choice is equivalent to choosing $\theta$ constant before integrating over the scalar. We shall return to this point in the following sections.

This example was intended to illustrate the general result that we will now present in the next section.

\section{Equivalence between original and enhanced versions of Abelian anomalous models}

Now, we return to the anomalous generic gauge model defined in (\ref{effective}), where 
\begin{equation}
I[\psi,\bar{\psi},A]=I_{M(ano)}[\psi,\bar{\psi},A]+I_{S}[A],
\end{equation}
with $I_{M(ano)}[\psi,\bar{\psi},A]$ being the anomalous matter action and $I_{S}[A]$ is the gauge-invariant free bosonic one.

The breakdown of local gauge invariance of the effective action (\ref{noninv_effect_act}) is usually referred to as current non-conservation. To understand this, we see that, since the effective action is not gauge-invariant, we may say that we do not have the \textit{Noether} identity $\partial_{\mu}\left(-\frac{1}{e}\frac{\delta W[A]}{\delta A_{\mu}(x)}\right)\equiv 0$, \textit{i.e.}, identically

\begin{equation}
\mathcal{A}\equiv \partial_{\mu}\left(-\frac{1}{e}\frac{\delta W[A]}{\delta A_{\mu}(x)}\right)\neq 0.
\label{anomaly}
\end{equation}
The quantity defined by (\ref{anomaly}) is an \textit{anomaly}. To understand the relation between this quantity and current divergence, we notice that

\begin{equation}
\partial_{\mu}\left(-\frac{1}{e}\frac{\delta W[A]}{\delta A_{\mu}(x)}\right)\exp(iW[A])=\int d\psi d\bar{\psi}\partial_{\mu}\left(-\frac{1}{e}\frac{\delta I[\psi,\bar{\psi},A]}{\delta A_{\mu}(x)}\right)\exp\left(iI\left[\psi, \bar{\psi},A\right]\right).
\end{equation}
Since $I_{S}[A^{\theta}]=I_{S}[A]$, we have
\begin{equation}
\partial_{\mu}\left(-\frac{1}{e}\frac{\delta I_{S}[A]}{\delta A_{\mu}(x)}\right)\equiv 0,
\label{noe_id}
\end{equation}
and, therefore, 
\begin{equation}
\int d\psi d\bar{\psi}\partial_{\mu}J^{\mu}\exp\left(iI\left[\psi, \bar{\psi},A\right]\right)=\mathcal{A}\exp(iW[A]), \label{ano1}
\end{equation}
where
\begin{equation}
J^{\mu}(x)\equiv -\frac{1}{e}\frac{\delta I_{M(ano)}[\psi,\bar{\psi},A]}{\delta A_{\mu}(x)}
\label{classical_curr}
\end{equation}
is the classical conserved current that may be obtained by global invariance of the action. If $\mathcal{A}$ is considered non-null, then eq. (\ref{ano1}) means current conservation breakdown at the quantum level, representing one of the most intriguing problems in quantum field theory. In this sense, to be very precise in our purposes, we \textit{define} the anomaly (\ref{ano1}), by generalizing it to the mean expectation value of the classical current divergence over the remaining fields besides the gauge one,

\begin{equation}
\int d\phi d\psi d\bar{\psi}\partial_{\mu}J^{\mu}\exp\left(iI\left[\psi, \bar{\psi},A, \phi\right]\right)=\mathcal{A}\exp(iW[A]), \label{ext_ano}
\end{equation}
where $\phi$ represents all other fields that may enter the theory beside the ones being considered, and an \textit{anomalous model} as being the one whose anomaly defined in (\ref{ext_ano}) is not \textit{identically} null.

Although such theories may bring theoretical problems, we may alternatively see an anomalous model as a faithful one, take the gauge field equation of motion from the effective action

\begin{equation}
\frac{\delta W[A]}{\delta A_{\mu}(x)}=0,
\end{equation}
and, in straight analogy with the Proca model, obtain the nullity of the anomaly as a \textit{subsidiary} condition

\begin{equation}
\mathcal{A}\equiv \partial_{\mu}\left(-\frac{1}{e} \frac{\delta W[A]}{\delta A_{\mu}}(x)\right)=0.
\end{equation}

However, this means that the theory has constraints, and one then has to prove that the model is internally consistent. In order to do so, we shall analyze a concrete example, which is the anomalous chiral Schwinger model \cite{JR, HT} whose action is

\begin{equation}
I_{sch}[\psi,\bar{\psi},A]=\int d^{x}\left\{-\frac{1}{4}F^{\mu\nu}F_{\mu\nu}+\bar{\psi}i\gamma^{\mu}[\partial_{\mu}-ieA_{\mu}P_{+}]\psi\right\},
\end{equation}
where
\begin{equation}
P_{+}\equiv \frac{1}{2}(1+\gamma_{5}).
\end{equation}
This action is gauge-invariant, and the classical conserved current obtained by its symmetry is given by

\begin{equation}
J^{\mu}=\bar{\psi}\gamma^{\mu}P_{+}\psi.
\end{equation}
The effective action is exactly soluble \cite{JR}, and given by

\begin{equation}
W_{sch}[A]=\int d^{2}x\left\{-\frac{1}{4}F^{\mu\nu}F_{\mu\nu}+\frac{e^{2}}{8\pi}A_{\mu}\left[ag^{\mu\nu}-(g^{\mu\alpha}+\epsilon^{\mu\alpha})\frac{\partial_{\alpha}\partial_{\beta}}{\square}(g^{\beta\nu}+\epsilon^{\beta\nu})\right]A_{\nu}\right\}, \label{schwinger}
\end{equation}
where $g^{\mu\nu}$ is the 2-dimensional Minkowski metric, $\epsilon^{\mu\alpha}$ is the Levi-Civita tensor and $a$ is an arbitrary regularisation parameter.

Now, it is easy to see that $W_{sch}[A^{\theta}]\neq W_{sch}[A]$ \cite{HT}. Indeed,

\begin{eqnarray}
\alpha_{1}[A,\theta]&=&W_{sch}[A^{\theta}]-W_{sch}[A]\nonumber\\
&=&\frac{1}{4\pi}\int d^{2}x\left\{\frac{1}{2}(a-1)\partial_{\mu}\theta\partial^{\mu}\theta-e\theta[(a-1)\partial_{\mu}A^{\mu}+\epsilon^{\mu\nu}\partial_{\mu}A_{\nu}]\right\}.
\label{Wess-Zumino}
\end{eqnarray}
Therefore, the chiral Schwinger model \textit{is} anomalous, with the anomaly being

\begin{equation}
\mathcal{A}=-\frac{e}{4\pi}\{(a-1)\partial_{\mu}A^{\mu}+\epsilon^{\mu\nu}\partial_{\mu}A_{\nu}\}.
\end{equation}
On the other hand, adopting the alternative point-of-view that we explained above, we may impose the variational principle to the effective action (\ref{schwinger}), and we just find the equation of motion of the vector field

\begin{equation}
\partial_{\mu}F^{\mu\nu}+\frac{e^{2}}{4\pi}\left(aA^{\nu}-\frac{\partial^{\nu}\partial^{\mu}}{\square}A_{\mu}+\epsilon^{\alpha\mu}\frac{\partial^{\nu}\partial_{\alpha}}{\square}A_{\mu}-\epsilon^{\nu\alpha}\frac{\partial_{\alpha}\partial^{\mu}}{\square}A_{\mu}+\epsilon^{\nu\alpha}\epsilon^{\beta\mu}\frac{\partial_{\alpha}\partial_{\beta}}{\square}A_{\mu}\right)=0. 
\label{mo_eq}
\end{equation} 
Taking the divergence of (\ref{mo_eq}) and using the fact that
\begin{equation}
\epsilon^{\mu\alpha}\epsilon^{\beta\nu}=g^{\mu\nu}g^{\alpha\beta}-g^{\mu\beta}g^{\alpha\nu},
\label{Levi-Civitta}
\end{equation}
we just arrive with the subsidiary condition that cancels the anomaly
\begin{equation}
(a-1)\partial_{\mu}A^{\mu}+\epsilon^{\mu\nu}\partial_{\mu}A_{\nu}=0.
\label{can_ano}
\end{equation}
Substituting it back to (\ref{mo_eq}), it is straightforward to find the Proca gauge-invariant version of the 2-dimensional vector field equation of motion
\begin{equation}
\partial_{\mu}F^{\mu\nu}+\frac{e^{2}}{4\pi}\frac{a^{2}}{(a-1)}\left(\eta^{\mu\nu}-\frac{\partial^{\nu}\partial_{\mu}}{\square}\right)A_{\mu}=0,
\label{mo_eq_Pro}
\end{equation}
but with the vector field restricted to the condition (\ref{can_ano}), instead of the \textit{Lorenz} gauge condition.

We now turn back to the general case and proceed the enhanced mapping. Using (\ref{inv-effect}), (\ref{enh_act_funct}), and (\ref{enh_act}), we just obtain

\begin{eqnarray}
\int &d\theta & d\psi d\bar{\psi} \partial_{\mu}\left(-\frac{1}{e}\frac{\delta[\psi, \bar{\psi}, A^{\theta}]}{\delta A_{\mu}(x)}\right)\exp\left(iI_{en}[\psi, \bar{\psi}, A, \theta]\right)\nonumber\\
=&\partial_{\mu} &\left(-\frac{1}{e}\frac{\delta W_{eff}[A]}{\delta A_{\mu}(x)}\right)\exp(iW_{eff}[A]).
\end{eqnarray}
Since $W_{eff}[A^{\theta}]=W_{eff}[A]$, we have the Noether identity

\begin{equation}
\partial_{\mu}\left(-\frac{1}{e}\frac{\delta W_{eff}[A]}{\delta A_{\mu}(x)}\right)\equiv 0,
\label{NI}
\end{equation}
and thus
\begin{equation}
\int d\theta d\psi d\bar{\psi} \partial_{\mu}\left(-\frac{1}{e}\frac{\delta I[\psi, \bar{\psi}, A^{\theta}]}{\delta A_{\mu}(x)}\right)\exp(iI_{en}[\psi, \bar{\psi}, A, \theta])\equiv 0.
\end{equation}
Since in fermonic theories the gauge fields are used to be coupled linearly to the matter fields, and the difference between $A_{\mu}$ and $A_{\mu}^{\theta}$ is just a translation, we may be sure that

\begin{equation}
\frac{\delta I_{M(ano)}[\psi, \bar{\psi},A^{\theta}]}{\delta A_{\mu}(x)}=\frac{\delta I_{M(ano)}[\psi, \bar{\psi},A]}{\delta A_{\mu}(x)}.
\end{equation}
By (\ref{noe_id}) we obtain, therefore

\begin{equation}
\int d\theta d\psi d\bar{\psi} \partial_{\mu}J^{\mu}\exp\left(iI_{en}[\psi, \bar{\psi}, A, \theta]\right)\equiv 0,
\end{equation}
which means that the Abelian enhanced formulation is anomaly-free.

As already discussed, the enhanced formulation, before the integration over the scalar, may be viewed as an anomalous analogue of the Stueckelberg mechanism \cite{eu}, and it clearly reduces to the original one by the gauge choice where $\theta$ is set constant. We now return to the example of the Chiral Schwinger model and get its enhanced version. Then we have, after integrating out the fermions,

\begin{equation}
W_{sch}[A^{\theta}]=\alpha_{1}[A,\theta]+W_{sch}[A].
\end{equation}
Therefore, we need only to consider the Wess-Zumino term (\ref{Wess-Zumino}) in the integration over $\theta$. Thus,
\begin{eqnarray}
&\exp&(iW_{eff}[A])\nonumber\\
&=&\int d\theta\exp\left(\frac{i}{4\pi}\int d^{2}x\left\{\frac{1}{2}(a-1)\partial_{\mu}\theta\partial^{\mu}\theta-e\theta[(a-1)\partial_{\mu}A^{\mu}+\epsilon^{\mu\nu}\partial_{\mu}A_{\nu}]\right\}\right). 
\label{Schwin1}
\end{eqnarray}
Using the fact that

\begin{eqnarray}
&\frac{i}{4\pi}&\int d^{2}x\left\{\frac{1}{2}(a-1)\partial_{\mu}\theta\partial^{\mu}\theta-e\theta[(a-1)\partial_{\mu}A^{\mu}+\epsilon^{\mu\nu}\partial_{\mu}A_{\nu}]\right\}\nonumber\\
&=&-\frac{i}{8\pi}(a-1)\int d^{x}\left[\frac{1}{\square}e\left(\partial_{\mu}A^{\mu}+\frac{1}{(a-1)}\epsilon{\mu\nu}\partial_{\mu}A_{\nu}\right)\right .\nonumber\\
&+&\left . \theta\right]\square\left[\frac{1}{\square}e\left(\partial_{\alpha}A^{\alpha}+\frac{1}{(a-1)}\epsilon^{\alpha\beta}\partial_{\alpha}A_{\beta}\right)+\theta\right]\nonumber\\
&-&\frac{e^{2}}{\square}\left(\partial_{\mu}A^{\mu}+\frac{1}{(a-1)}\epsilon^{\mu\nu}\partial_{\mu}A_{\nu}\right)\left(\partial_{\alpha}A^{\alpha}+\frac{1}{(a-1)}\epsilon^{\alpha\beta}\partial_{\alpha}A_{\beta}\right),
\end{eqnarray}
doing the following translation over the field $\theta$:
\begin{equation}
\theta '=\theta+\frac{1}{\square}e\left(\partial_{\mu}A^{\mu}+\frac{1}{(a-1)}\epsilon^{\mu\nu}\partial_{\mu}A_{\nu}\right),\qquad\qquad d\theta '=d\theta,
\label{SC}
\end{equation}
and proceeding the integration over the new parameter $\theta '$ in (\ref{Schwin1}), it is straightforward to find
\begin{eqnarray}
\exp(iW_{eff}[A])&=\exp(iW_{sch}[A])\int d\theta \exp\left\{i\frac{e^{2}}{8\pi}(a-1)\int d^{2}x\left(\partial_{\mu}A^{\mu}+\frac{1}{(a-1)}\epsilon_{\mu\nu}\partial_{\mu}A_{\nu}\right)\right .\nonumber\\
&\left . \frac{1}{\square}\left(\partial_{\mu}A^{\mu}+\frac{1}{(a-1)}\epsilon_{\mu\nu}\partial_{\mu}A_{\nu}\right)\right\}.
\end{eqnarray}
Using (\ref{schwinger}), and (\ref{Levi-Civitta}), we finally obtain
\begin{equation}
W_{eff}[A]=\int d^{2}x\left\{-\frac{1}{4}F^{\mu\nu}F_{\mu\nu}+\frac{1}{2}\frac{e^{2}}{4\pi}\frac{a^{2}}{(a-1)}A_{\mu}\left[g^{\mu\nu}-\frac{\partial^{\mu}\partial^{\nu}}{\square}\right]A_{\nu}\right\}.
\label{Proca_Inv}
\end{equation}
This result was also found in ref. \cite{HT}. We may observe that the effective action (\ref{Proca_Inv}) is exactly the Proca 2-dimensional gauge-invariant action that gives the equation of the anomalous original model (\ref{mo_eq_Pro}), but without the restriction (\ref{can_ano}) over the gauge field. Therefore, analogously to the \textit{Proca/Stueckelberg} case, if we fix the gauge by (\ref{can_ano}) in the gauge-invariant effective anomalous action, then the enhanced model reduces to the original anomalous one, showing equivalence between both formulations.

\section{Discussion}

The examples mentioned above may lead us to a very simple and clear conclusion, which is that a gauge theory is equivalent to a non-gauge theory if the first is reducible to the second one by a gauge choice. By virtue of Pauli's condition, the original and enhanced formulations are straightforwardly equivalent, since the second reduces to the first one by the gauge choice where the Stueckelberg scalar $\theta$ is set to be constant.

From the canonical gauge theory point-of-view, on the other hand, our examples show that the integrated effective models are reducible to one-another by the Lorenz gauge choice
\begin{equation}
\partial_{\mu}A^{\mu}=0
\label{PG2}
\end{equation}
in the Proca case, and the rather distinct one
\begin{equation}
(a-1)\partial_{\mu}A^{\mu}+\epsilon^{\mu\nu}\partial_{\mu}A_{\nu}=0
\label{SG3}
\end{equation}
in the chiral Schwinger model. We see that, to achieve these gauge conditions, we have to do the following transformations over a not restricted gauge field $A_{\mu}$:
\begin{equation}
A'_{\mu}=A_{\mu}+\frac{1}{e}\partial_{\mu}\Lambda
\label{GT}
\end{equation}
taking the divergence of $A'_{\mu}$ in the Proca case in (\ref{GT}), we have
\begin{equation}
\partial_{\mu}A'^{\mu}=\partial_{\mu}A^{\mu}+\frac{1}{e}\square\Lambda_{P}=0
\label{SG}
\end{equation}
which means that
\begin{equation}
\Lambda_{P}=-\frac{e}{\square}\partial_{\mu}A^{\mu}.
\label{PG}
\end{equation}
Doing the same for the chiral Schwinger model, and adding $\frac{1}{(a-1)}\epsilon^{\mu\nu}\partial_{\mu}A_{\nu}$, it is straightforward to find
\begin{equation}
\Lambda_{sch}=-\frac{e}{\square}\left(\partial_{\mu}A^{\mu}+\frac{1}{(a-1)}\epsilon^{\mu\nu}\partial_{\mu}A_{\nu}\right).
\label{SG2}
\end{equation}
If we compare (\ref{PG}) and (\ref{SG2}) with (\ref{PC}) and (\ref{SC}), respectively, we see that the translation over the field $\theta$ to reach the pure gauge invariant action is just
\begin{equation}
\theta\rightarrow\theta '=\theta-\Lambda.
\end{equation}
This suggests that the enhanced gauge condition $\theta (x)=constant$ that ensures equivalence between both models is transferred to the gauge fields after integrated out the Stueckelberg, as manifested in (\ref{PG2}) and (\ref{SG3}), in such a way that it turns to be the subsidiary conditions of the original models.

We now turn our attention to the standard formulation. The work of ref. \cite{LRR}, in particularly analyzing the standard version of the chiral Schwinger model, shows that its gauge-invariant correlation functions coincide with those of the original anomalous theory, but it also shows that it is not the case for gauge dependent Green's functions, and no choice of gauge conditions exists for which the generating functional of the standard formulation coincides with that of the original theory. The conclusion is, thus, that its physical contents are quite different. However, it was shown that the action with with the addition of the Wess-Zumino term is equivalent to the original anomalous if the gauge condition (\ref{SG3}) is \textit{imposed} in both models. As it was shown, this condition may arise from the original model as a subsidiary condition. On the other hand, besides the final effective action is made gauge-invariant, the initial one is not, since the Wess-Zumino term breaks it. To understand what it means, we consider a model with the standard action (\ref{standard}) and try to obtain the conserved quantity given by the gauge invariance of the effective action, we then find
\begin{eqnarray}
&\partial_{\mu}&\left(-\frac{1}{e}\frac{\delta W_{eff}[A]}{\delta A_{\mu}(x)}\right)\exp(iW_{eff}[A])\nonumber\\
&=&\int d\theta d\psi d\bar{\psi}\partial_{\mu}\left(-\frac{1}{e}\frac{\delta I_{st}[\psi, \bar{\psi}, A, \theta}{\delta A_{\mu}(x)}\right)\exp(iI_{st}[\psi, \bar{\psi}, A, \theta])=0
\end{eqnarray}
or, by (\ref{standard}) and (\ref{classical_curr})
\begin{equation}
\int d\theta d\psi d\bar{\psi} \partial_{\mu}J^{\mu}\exp (iI_{st}[\psi,\bar{\psi},A,\theta])=\int d\theta d\psi d\bar{\psi}\partial_{\mu}\left(\frac{1}{e}\frac{\delta\alpha_{1}[A,\theta]}{\delta A_{\mu}(x)}\right)\exp(iI_{st}[\psi,\bar{\psi},A,\theta]).
\label{ano2}
\end{equation}
If we integrate the right hand side of (\ref{ano2}) and use (\ref{NI}), we will just obtain

\begin{equation}
\int d\theta d\psi d\bar{\psi} \partial_{\mu}J^{\mu}\exp (iI_{st}[\psi,\bar{\psi},A,\theta])=\mathcal{A}\exp(iW_{eff}[A]),
\label{mean_curr}
\end{equation}
with $\mathcal{A}$ given by (\ref{anomaly}), which means, by our definition (\ref{ext_ano}), that the standard formulation is still anomalous. we can notice that, in this model, unlike the original anomalous one, no subsidiary condition arises in order to cancel the anomaly. To be more precise, a kind of subsidiary condition arises if we use the Dyson-Schwinger equation for the field $\theta$. To see this, we write

\begin{eqnarray}
&\int& d\theta d\psi d\bar{\psi} \frac{\delta I_{st}}{\delta{\theta}}\exp(iI_{st}[\psi,\bar{\psi},A,\theta])\nonumber\\
&=&\int d\theta d\psi\frac{\delta \alpha_{1}}{\delta\theta}[A,\theta]\exp(iI_{st}[\psi,\bar{\psi},A,\theta])=0,
\end{eqnarray}
but
\begin{eqnarray}
\frac{\delta \alpha_{1}}{\delta\theta}[A,\theta]=\frac{\delta W}{\delta\theta}[A^{\theta}]&=&\int d^{n}y\frac{\delta W[A^{\theta}]}{\delta A_{\mu}^{\theta}(y)}\frac{\delta A_{\mu}^{\theta}(y)}{\delta\theta(x)}\nonumber\\
&=&\int d^{n}y\frac{1}{e}\frac{\delta W[A^{\theta}]}{\delta A_{\mu}^{\theta}(y)}\partial_{\mu}\delta (x-y)=\partial_{\mu}\left(-\frac{1}{e}\frac{\delta W[A^{\theta}]}{\delta A^{\theta}_{\mu}(x)}\right)=\mathcal{A}^{\theta}
\end{eqnarray}
and, therefore
\begin{equation}
\int d\theta d\psi d\bar{\psi} \mathcal{A}^{\theta}\exp(iI_{st}[\psi,\bar{\psi},A,\theta])=0.
\label{mean_transf_ano}
\end{equation}

We see that, if the anomaly is made gauge-invariant, which means to set $a=1$ in the chiral Schwinger model, then the left hand side of (\ref{mean_transf_ano}) reduces to (\ref{mean_curr}) and the anomaly cancels out. However, in our simplest example, the choice $a=1$ represents a gauge parameter (\ref{SG2}), to be used in order to integrate the scalar by translation in (\ref{SC}), which is infinite. It is easy to see, by eq. (\ref{Schwin1}), that such a choice represents a functional Dirac delta that has the anomaly as its parameter, that is, if $a=1$, then
\begin{eqnarray}
\exp(iW_{eff}[A])&=&\exp(iW_{sch}[A])\int d\theta\exp\left(-\frac{i}{4\pi}\int d^{2}x\{e\theta\epsilon^{\mu\nu}\partial_{\mu}A_{\nu}\}\right)\nonumber\\
&=&\delta (\mathcal{A}[A])\exp (iW_{sch}[A]).
\end{eqnarray}
This is, indeed, trivially equivalent to the original anomalous model, since it has just the same non-gauge action with the redundant vanishing anomaly condition being imposed before the equation of motion of the vector field is taken from the effective action. The surviving anomaly in (\ref{mean_curr}) is, thus, caused by its gauge non-invariance. On the other hand, another value of $a$ clearly turns the anomaly cancelation impossible.

Yet, we see that if $a\neq 1$, the anomaly will not be invariant, but the final effective action will be. In this sense, if we were allowed to choose a gauge condition such as $\mathcal{A}=0$, then the anomaly would cancel out, the current would become conserved and the standard formulation would turn to be the original one. We may also see that the standard formulation reduces to the original one if we set $\theta (x)=constant$. However, evidently the situation imposed by such condition is not physically equivalent to leaving the anomaly non-null, since we have no current conservation in one situation, and have it conserved in another one. Therefore, we have a breaking of the physical equivalence between distinct gauge configurations in the final gauge-invariant effective action of the standard formulation due to the fact that the standard action is not gauge-invariant. These considerations may explain the results found in ref. \cite{LRR}.

On the other hand, we may try to adopt the point-of-view in which there may be a modified conserved current on the standard formulation due to the addition of the Wess-Zumino term. To calculate such current, we need to use the gauge invariance of the final effective action, which gives us, from (\ref{ano2})
\begin{equation}
\int d\theta d\psi d\bar{\psi}\partial_{\mu}\left(J^{\mu}-\frac{1}{e}\frac{\delta\alpha_{1}[A,\theta]}{\delta A_{\mu}(x)}\right)\exp(iI_{st}[\psi,\bar{\psi},A,\theta])\equiv 0.
\end{equation}
Therefore, the standard current must be taken from
\begin{equation}
J^{mu}_{st}=J^{\mu}-\frac{1}{e}\frac{\delta\alpha_{1}[A,\theta]}{\delta A_{\mu}}(x).
\label{modif_curr}
\end{equation} 
However, as it was seen in (\ref{mean_curr})

\begin{equation}
\int d\theta d\psi d\bar{\psi} \partial_{\mu}J^{\mu}\exp (iI_{st}[\psi,\bar{\psi},A,\theta])=\mathcal{A}\exp(iW_{eff}[A]). \nonumber
\end{equation}
If we choose our gauge choice where the anomaly is cancelled out $\mathcal{A}=0$, we arrive at the same classical current
\begin{equation}
J^{\mu}_{st}=J^{\mu}
\end{equation}
in one specific gauge, and such modified one (\ref{modif_curr}) in all others, showing that such current cannot be physical. This may be explained, once again, by the non-invariance of the standard action.

It should be noticed, though, that such difference between the standard and enhanced models appears only before the integration over the matter fields. After that, the same intermediary effective action (the one containing $A$ and $\theta$) is found. In the sense of the effective theory, thus, this means that there is no net difference whether we are working initially with one or other formulation.

\section{Correspondence between the gauge-invariant chiral Schwinger and the two-dimensional Stueckelberg models}

We saw that both gauge-invariant formulations of the  chiral Schwinger model give us, after integrated the fermions
\begin{equation}
W_{sch}[A^{\theta}]=W_{sch}[A]+\int d^{2}x\left\{\frac{1}{8\pi}(a-1)\partial_{\mu}\theta\partial^{\mu}\theta-e\theta[(a-1)\partial_{\mu}A^{\mu}+\epsilon^{\mu\nu}\partial_{\mu}A_{\nu}]\right\},
\end{equation}
while the Stueckelberg gauge-invariant model is described by
\begin{equation}
W_{P}[A^{\theta}]=-\frac{1}{4}\int d^{4}xF^{\mu\nu}F_{\mu\nu}+\frac{m^{2}}{2}\int d^{4}x\left(A^{\mu}+\frac{1}{e}\partial^{\mu}\theta\right)\left(A_{\mu}+\frac{1}{e}\partial_{\mu}\theta\right).
\end{equation}
All these models, after integrated out the scalar, reach the same gauge-invariant Proca action
\begin{equation}
W_{eff}[A]=\int d^{2}x\left\{-\frac{1}{4}F^{\mu\nu}F_{\mu\nu}+\frac{1}{2}\frac{e^{2}}{4\pi}\frac{a^{2}}{(a-1)}A_{\mu}\left[g^{\mu\nu}-\frac{\partial^{\mu}\partial^{\nu}}{\square}\right]A_{\nu}\right\}.
\end{equation}
The difference between them relies on the translation over the $\theta$ variable. While in the Stueckelberg model we change variables such as
\begin{equation}
\theta_{P}\rightarrow \theta_{P}+\frac{e}{\square}\partial_{\mu}A^{\mu}
\end{equation}
in the chiral Schwinger one, we have
\begin{equation}
\theta_{sch}\rightarrow \theta_{sch}+\frac{e}{\square}\left(\partial_{\mu}A^{\mu}+\frac{1}{(a-1)}\epsilon^{\mu\nu}\partial_{\mu}A_{\nu}\right).
\end{equation}
Therefore, it is very convenient to find a map between these models. Indeed, it is straightforward to check that the relation
\begin{equation}
\theta_{sch}=\frac{a}{(a-1)}\theta_{P}-\frac{e}{(a-1)}\frac{1}{\square}\epsilon^{\mu\nu}\partial_{\mu}A^{\nu}+\frac{e}{(a-1)}\frac{1}{\square}\partial_{\mu}A^{\mu}
\end{equation}
turns one model to the other
\begin{eqnarray}
W_{sch(en)}[A,\theta_{sch}]&=&W_{P(en)}[A,\theta_{P}]\\
\int d\theta_{sch}\exp(iW_{sch(en)}[A,\theta_{sch}])&\sim&\int d\theta_{P}\exp(iW_{P(en)}[A,\theta_{P}]),
\end{eqnarray}
which means that the gauge-invariant chiral Schwinger model, after integrated out the fermions, may be identified with the original Stueckelberg model, which is known to be unitary and renormalizable in some gauge fixing condition \cite{LS, vanHees}.

\section{Conclusion}
This work illustrates a simple yet controversial idea: that a gauge-invariant model may be equivalent to a non gauge-invariant one. This is possible, as it was shown, as long as one is reducible to the other one by some gauge condition. The controverse relying on this result might be related to the idea that current conservation is due to gauge symmetry. However, at least in the classical case, Noether's theorem ensures current conservation through \textit{global} gauge invariance and the variational principle, instead of a rather strong condition, which is local gauge symmetry, as it becomes evident in the Proca case. Work is in progress to clarify the relation between local gauge symmetry and current conservation in the context of quantum models.

On the other hand, this idea becomes manifest by an interesting procedure of recovering gauge symmetry from non-gauge theories, and the simplest example of the chiral Schwinger model shows us, after integrating out the fermions, that the effective theory may be identified with the original theory proposed by Stueckelberg, showing a possible alternative mechanism of mass generation from chiral fermions that might be unitary and renormalizable.

Finally, we may suppose that such a mass mechanism from chiral fermions is valid for higher dimensions. If one can prove that the final effective theory $W_{eff}[A]$ is the Proca invariant one (\ref{Proca_Inv}) even in 4 dimensions, then it will be given more than the possibility of quantisation of abelian anomalous models, but its connection with a quantizable mass generation mechanism able to substitute the Higgs, ate least for the abelian case.

\begin{acknowledgments}
I am very grateful to Professor Abhay Ashtekar and the \textit{Institute for Gravitation and the Cosmos} (Penn State University) for the hospitality, and Dr. Marc Geiller for proof reading. This work was financially supported by CEFET, through the CODIB (Coordena\c{c}\~{a}o de Disciplinas B\'{a}sicas) from Uned, Nova Igua\c{c}u, RJ, Brazil, and in part by the NSF grant PHY 1205388 and the Eberly research funds of Penn State University.
\end{acknowledgments}

\end{document}